\documentclass[runningheads]{llncs}
\usepackage[T1]{fontenc}
\usepackage{graphicx,verbatim, hyperref}
\usepackage[acronym]{glossaries}
\usepackage{booktabs}
\usepackage{makecell}
\usepackage{color}
\usepackage{orcidlink}

\usepackage{amsmath,amsfonts,bm}

\def\1{\bm{1}}

\def\rvepsilon{{\mathbf{\epsilon}}}

\def\rvh{{\mathbf{h}}}

\def\rvp{{\mathbf{p}}}

\def\rvx{{\mathbf{x}}}
\def\rvy{{\mathbf{y}}}

\def\rmI{{\mathbf{I}}}

\def\mSigma{{\bm{\Sigma}}}

\DeclareMathAlphabet{\mathsfit}{\encodingdefault}{\sfdefault}{m}{sl}
\SetMathAlphabet{\mathsfit}{bold}{\encodingdefault}{\sfdefault}{bx}{n}

\def\gN{{\mathcal{N}}}

\def\sR{{\mathbb{R}}}

\newcommand{\E}{\mathbb{E}}

\DeclareMathOperator{\diag}{diag}

\begin{document}

\title{Semantic Diffusion Posterior Sampling for Cardiac Ultrasound Dehazing}

\author{Tristan S.W. Stevens \orcidlink{0000-0002-8563-59315} \and
Oisín Nolan \orcidlink{0009-0002-6939-7627} \and
Ruud J.G. van Sloun \orcidlink{0000-0003-2845-0495}}

\authorrunning{Stevens et al.}

\institute{Eindhoven University of Technology, the Netherlands
\email{\{t.s.w.stevens,o.i.nolan,r.j.g.v.sloun\}@tue.nl}}

\newacronym{DPS}{DPS}{Diffusion Posterior Sampling}
\newacronym{dm}{DM}{diffusion model}
\newacronym{dgm}{DGM}{deep generative model}
\newacronym{gan}{GAN}{Generative Adversarial Network}
\newacronym{fid}{FID}{Fréchet Inception Distance}
\newacronym{ks}{KS}{Kolmogorov–Smirnov}
\newacronym{asd}{ASD}{Average Surface Distance}
\newacronym{kid}{KID}{Kernel Inception Distance}
\newacronym{dsm}{DSM}{denoising score matching}

\maketitle

\begin{abstract}
Echocardiography plays a central role in cardiac imaging, offering dynamic views of the heart that are essential for diagnosis and monitoring. However, image quality can be significantly degraded by haze arising from multipath reverberations, particularly in difficult-to-image patients. In this work, we propose a semantic-guided, diffusion-based dehazing algorithm developed for the MICCAI Dehazing Echocardiography Challenge (DehazingEcho2025). Our method integrates a pixel-wise noise model, derived from semantic segmentation of hazy inputs into a diffusion posterior sampling framework guided by a generative prior trained on clean ultrasound data. Quantitative evaluation on the challenge dataset demonstrates strong performance across contrast and fidelity  metrics. Code for the submitted algorithm is available on GitHub.\footnote{\url{https://github.com/tristan-deep/semantic-diffusion-echo-dehazing}\label{code}}.


\end{abstract}

\section{Introduction}
Ultrasound is a popular modality for cardiac imaging due to its high temporal resolution, cost effectiveness, and real-time imaging capabilities, enabling the detection of a variety of cardiac abnormalities \cite{barry2023role}. An ongoing challenge in echocardiography is that of clutter or haze resulting from multipath reverberations \cite{sjoerdsma2020spatial,stevens2024dehazing}, which can prevent accurate measurement from B-Mode images. This has motivated the development \textit{dehazing algorithms}, which aim to recover clean images $\rvx$ from hazy input images $\rvy$.

Recently, a number of dehazing algorithms leveraging \acrfullpl{dgm} have been proposed, using prior knowledge of the clean image distribution to infer sets of clean images corresponding to observed hazy images. One such approach involves using \acrfullpl{gan} to perform domain adaptation, wherein the style of one dataset is transferred to samples from another \cite{xia2022multilevel,huang2022stability} while retaining structural contents. Other approaches opt for \acrfullpl{dm}~\cite{ho2020denoising,song2020score}, which are known to represent the state-of-the-art in image synthesis \cite{dhariwal2021diffusion}. One such method, introduced by Stevens \textit{et al.}, involves using \acrshortpl{dm} to learn models of the distributions of both clean images and haze, which are then used to separate the clean tissue and haze components of the input hazy image \cite{stevens2024dehazing}. However, in this work, the signal model employed is defined on pre-envelope-detected signals, which are in some cases not available. This motivates the development of diffusion-based dehazing algorithms that operate in the image domain. In this paper, we propose such an algorithm, \textit{Semantic Diffusion Posterior Sampling}, which first computes a semantic segmentation map estimating the haze content of each pixel, and then uses the \acrfull{DPS} algorithm to generate posterior samples from the clean image distribution given the hazy measurement. The estimated haze map serves to control the strength of the conditional guidance during the image generation process, according more with the measurements in clean regions, and less in hazy regions.

\section{Challenge}\label{sec:challenge}
This work was developed in the context of the MICCAI Dehazing Echocardiography Challenge (DehazingEcho2025)\footnote{\url{https://dehazingecho2025.grand-challenge.org/}}, which aims to enhance the quality of transthoracic echocardiographic images acquired from difficult-to-image patients. The dataset provided for this challenge comprises two subsets: (1) a clean set of 4,376 frames obtained from 75 easy-to-image subjects, and (2) a noisy set of 2,324 frames acquired from 40 difficult-to-image subjects. Each image in the dataset is part of a 60-frame four-chamber view cine-loop.

For quantitative benchmarking, a hidden test set of 536 noisy frames is used for online evaluation on the Grand Challenge platform~\cite{challenge2021grand}. The evaluation protocol incorporates multiple complementary metrics designed to capture different aspects of image quality and utility. Specifically:
\begin{itemize}
\item \textbf{\acrfull{fid}} assesses perceptual similarity between denoised and clean images.
\item \textbf{CNR} and \textbf{gCNR} quantify contrast between myocardial tissue (septum) and noise-prone regions (left ventricular).
\item \textbf{\acrfull{ks}} test measures distributional similarity to assess structure preservation of the septum and noise removal in the ventrical.
\item \textbf{Dice coefficient} and \textbf{\acrfull{asd}} evaluate the compatibility of denoised images with downstream segmentation tasks, using a pre-trained universal ultrasound foundation model (USFM)~\cite{jiao2024usfm} targeting the left ventricle.
\end{itemize}
The \emph{final score} is derived from a weighted aggregation of the above metrics, balancing denoising performance, structural preservation, and downstream task impact with respective weights of 5:3:2. 

\section{Method}
The algorithm consists of two primary steps: first, generating a semantic segmentation mask based on the input hazy image, and second, using that mask to guide a diffusion model towards generating a dehazed image. We frame the task of dehazing as a Bayesian inverse problem with the following forward model:

\begin{equation}
    \rvy = \rvx + \rvh, \qquad \rvh\sim\gN(0,\mSigma), \quad \rvx\sim p_{\text{generative model}}(\rvx)
\label{eq:forward-model}
\end{equation}
where $\rvy, \rvx, \rvh \in \sR^n$ denote the hazy measurement, clean image, and residual haze respectively. The haze is modeled as additive, zero-mean Gaussian noise with a spatially varying diagonal covariance matrix:
\begin{align}
\mSigma^{-1} =\diag(\sigma_1^{-2}, \sigma_2^{-2}, \ldots, \sigma_n^{-2})=\diag(\rvp),
\end{align}
where each variance $\sigma_i^2$ is determined by the pixel-wise semantic segmentation map $\rvp$ as outlined in Section~\ref{sec:segmentation}. To solve \eqref{eq:forward-model}, we employ a diffusion prior, which we detail in the following section.

\subsection{Deep generative prior for clean cardiac ultrasound}
We start with leveraging a \acrfull{dgm} to model the distribution $p(\rvx)$ of clean echocardiography images $\rvx$. Specifically, we train a \acrfull{dm}, and subsequently perform posterior sampling $p(\rvx | \rvy)$ conditioned on the hazy inputs $\rvy$. \\

\noindent\textbf{Training:} Diffusion models learn to approximate the gradient of the log-density (i.e., the score function) of the data distribution by denoising progressively noised inputs. The training objective is based on \acrfull{dsm}, which seeks to minimize the discrepancy between predicted and true noise across different noise levels:
\begin{align}
    \mathcal{L}_{\text{DSM}}(\theta) = \E_{\rvx_0\sim p(\rvx_0), \rvepsilon \sim \mathcal{N}(\mathbf{0}, \rmI), \tau \sim \mathcal{U}(0,\mathcal{T})} \left[ \left\| \rvepsilon_\theta(\rvx_\tau, \tau) - \rvepsilon \right\|^2 \right],
\label{eq:dsm}
\end{align}
where $\rvx_\tau = \alpha_\tau \rvx_0 + \sigma_\tau \rvepsilon$ denotes a corrupted version of a sample from our clean dataset $\rvx_0$ at a continuous noise level $\tau$, with a pre-defined noise schedule, parameterized by $\alpha_\tau$ and $\sigma_\tau$. The model $\rvepsilon_\theta(\rvx_\tau, \tau)$ is trained to predict the noise component $\rvepsilon \sim\mathcal{N}(\mathbf{0}, \rmI)$ from the corrupted input $\rvx_\tau$, effectively learning the score $\rvepsilon_\theta(\rvx_\tau, \tau)\approx -\sigma_\tau\nabla_{\rvx_\tau} \log p(\rvx_\tau)$. To further improve the perceptual quality of generated samples (note the \acrshort{fid} objective of the challenge in Section~\ref{sec:challenge}), we additionally incorporate a \acrfull{kid} loss~\cite{binkowski2018demystifying}, computed between generated and real clean images. Unlike \acrshort{fid}, \acrshort{kid} is unbiased and better suited for comparing small datasets. Formally, the \acrshort{kid} loss is defined as the squared maximum mean discrepancy between the feature embeddings of a pretrained InceptionV3 network of generated and real samples. This encourages the model to produce samples that align more closely with the distribution of clean dataset. The total training loss is then given by:
\begin{align}
    \mathcal{L}(\theta) = \mathcal{L}_{\text{DSM}}(\theta) + \lambda_{\text{KID}} \cdot \mathcal{L}_{\text{KID}}(\theta),
\label{eq:total_loss}
\end{align}
where $\lambda_{\text{KID}}$ is a weighting factor controlling the influence of the perceptual loss. The diffusion model is pre-trained on the publicly available EchoNet-LVH dataset~\cite{duffy2022high}, and finetuned on the clean partition of the DehazingEcho2025 challenge dataset. For more details see Table~\ref{tab:model_summary}.

\subsection{Semantic Segmentation}\label{sec:segmentation}
The first step of the algorithm involves generating a segmentation mask from the hazy image which estimates the haze content of each pixel. This segmentation mask provides a \textit{noise level} for each pixel, defining a forward model for \acrfull{DPS}~\cite{chung2022diffusion} wherein high-signal pixels provide strong guidance, closely matching the measured pixels, and high-noise pixels provide weak guidance, falling back towards the prior distribution on dehazed images.

\begin{figure}
    \centering
    \includegraphics[width=1\linewidth]{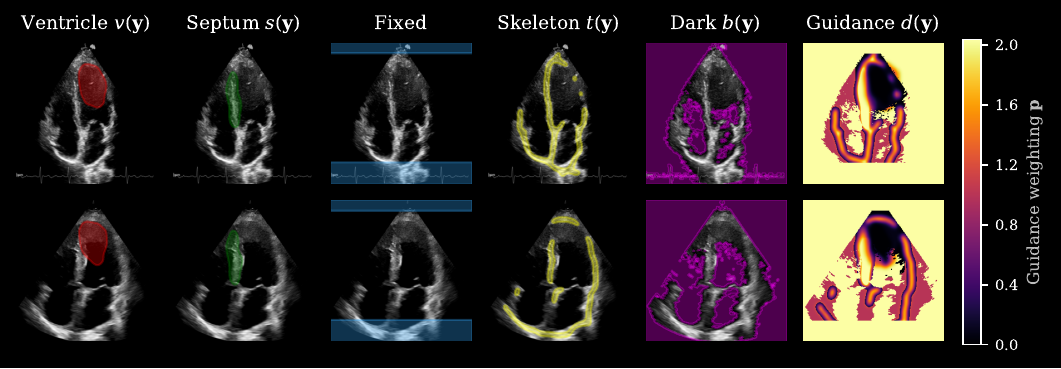}
    \caption{Visualization of the individual components for the semantic segmentation and the resulting constructed guidance weighting map $\rvp$ for two different patients.}
    \label{fig:segmentation}
\end{figure}

In order to construct these semantic segmentation masks, a combination of learned and classical segmentation methods was used.
\begin{itemize}
    \item \textbf{Ventricle and Septum Segmentation}: In order to identify the ventricles and septum, a DeepLabV3+ \cite{peng2020semantic} model was trained on manually annotated regions of interest corresponding to the septum and ventricle, provided in the challenge dataset. We denote the resulting masks as $v(\rvy)$ and $s(\rvy)$, identifying the ventricle and septum from the hazy image $\rvy$. The DeepLabV3+ model outputs a map of logits, which are thresholded by a parameter $\theta$ to create a binary mask, which is then blurred using a Gaussian kernel with standard deviation $\sigma_\text{blur}$.
    \item \textbf{Tissue Segmentation}: In order segment regions of tissue, the hazy image was skeletonized using \texttt{skimage.morphology.skeletonize}\footnote{\url{https://scikit-image.org/docs/0.25.x/api/skimage.morphology.html\#skimage.morphology.skeletonize}} \cite{van2014scikit}, a function which implements the thinning algorithm proposed by Zhang \textit{et al.} \cite{zhang1984fast}, mapping the hazy input image to thin skeleton tracing the tissue. The skeleton mask is denoted $t(\rvy)$.
    \item \textbf{Fixed pixels}: Two bands of pixels, at the top and bottom of the image, are kept fixed to ensure the preservation of details from the DICOM overlay.
    \item \textbf{Background pixels}: Any pixel not present in $v(\rvy)$, $s(\rvy)$, or $t(\rvy)$ is considered a \textit{background pixel}, $b(\rvy)$.
    \item \textbf{Dark pixels}: Any pixel for which $\rvy < 1e^{-6}$ is segmented as a \textit{dark pixel}, $d(\rvy)$.
\end{itemize}
The final mask is then created by taking a weighted sum of these individual masks, producing the precision vector $\rvp$ populating the diagonal of the precision  matrix $\mSigma^{-1}$ used in the \acrshort{DPS} forward model:
\begin{equation}
    \rvp = \omega b(\rvy) + \omega_v v(\rvy) + \omega_s (s(\rvy) + t(\rvy) + d(\rvy)) 
\end{equation}
Finally, we handle the edge case where the skeleton $t(\rvy)$ crosses the ventricle $v(\rvy)$, splitting it in two, by setting $\omega_v = 0$. An overview of the individual segmentation steps, and the resulting guidance weighting map is shown in Fig.~\ref{fig:segmentation}. Notably, imperfections introduced at one stage, such as over-segmentation or missed structures, can often be mitigated by complementary information from other segmentation steps, resulting in a more robust pixel-wise noise estimation.

\subsection{Semantic Diffusion Posterior Sampling}
Sampling from the posterior distribution $\rvx_0\sim p(\rvx_0 | \rvy)$ is achieved by initiating $\rvx_\mathcal{T}$ as a Gaussian random vector, and then applying an iterative denoising process that progressively refines the sample across decreasing noise levels as follows:
\begin{enumerate}
    \item Estimate the clean image from the current corrupted input:
    \begin{align}
        \hat{\rvx}_0 \leftarrow \frac{1}{\alpha_{\tau}} \left( \rvx_{\tau} - \sigma_{\tau} \rvepsilon_\theta(\rvx_{\tau}, \tau) \right).
    \label{eq:ddim_x0}
    \end{align}
    \item Guide the sample towards our hazy measurement with the forward model defined in \eqref{eq:forward-model}, along with a penalty on a smoothed L1 norm\footnote{\url{https://docs.pytorch.org/docs/stable/generated/torch.nn.SmoothL1Loss.html}}, parameterized by $\beta$, on the pixels in the ventricle.
    \begin{align}
        \hat{\rvx}_0 \leftarrow \hat{\rvx}_0 - \frac{1}{2} \nabla_{\rvx_\tau}(\rvy - \hat{\rvx}_0)^\top\mSigma^{-1}(\rvy - \hat{\rvx}_0) - \eta \nabla_{\rvx_\tau}|v(\rvy)\odot \hat{\rvx}_0|_\beta.
    \end{align}
    \item Predict the next corrupted sample at a lower noise level $\tau' < \tau$:
    \begin{align}
        \rvx_{\tau'} \leftarrow \alpha_{\tau'} \hat{\rvx}_0 + \sigma_{\tau'} \rvepsilon_\theta(\rvx_{\tau}, \tau).
    \label{eq:ddim_xt}
    \end{align}
\end{enumerate}
This deterministic denoising process is repeated until $\tau' = 0$, yielding a sample from the clean data distribution, conditioned on hazy measurement $\rvy$.

\subsection{Algorithm details}
\begin{figure}
    \centering
    \includegraphics[width=0.6\linewidth]{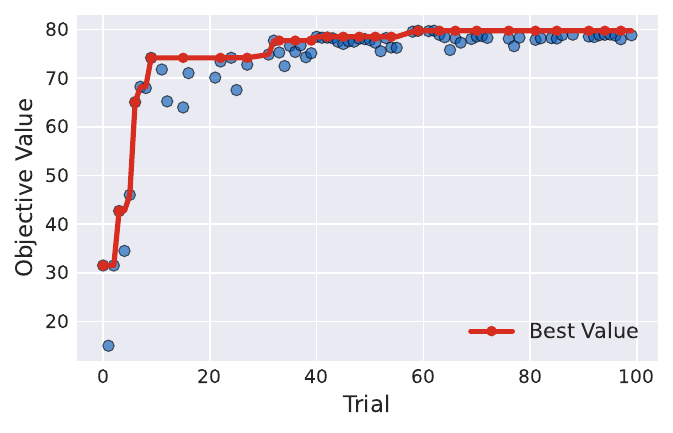}
    \caption{Hyperparameter optimization of 100 trials for the inference parameters listed in Table~\ref{tab:model_summary}, with the challenge's final score as objective.}
    \label{fig:optimization}
\end{figure}

A summary of the most important parameters used in each component of the dehazing algorithm is listed in Table~\ref{tab:model_summary}. Hyperparameters were optimized with \texttt{Optuna}~\cite{akiba2019optuna} using a subset of 237 images from the available noisy set that have corresponding masks. The optimization objective was set to the final score as specified in Section~\ref{sec:challenge}. A plot of the hyperparameter sweep is shown in Fig.~\ref{fig:optimization}. The final scores are slightly underestimated due to the \acrshort{fid}'s sensitivity to the smaller sample size, approximately $2\times$ lower than the online challenge test set.

The algorithm is implemented using \texttt{Keras 3} with the \texttt{JAX} backend for accelerated inference through JIT compilation. Furthermore, our implementation heavily relies on \texttt{zea}, a toolbox for cognitive ultrasound imaging~\cite{zea2025}\footnote{\url{https://zea.readthedocs.io/}}. Code with the complete algorithm implementation is available on GitHub~\footref{code}.

\begin{table}[t]
\centering
\caption{Overview of model components, architectures, training/inference settings, and datasets used.}
\label{tab:model_summary}
\begin{tabular}{@{}p{2.1cm}p{2.3cm}p{2.2cm}p{2.1cm}p{3.4cm}@{}}
\toprule
\textbf{Model} & \textbf{Architecture} & \textbf{Inference} & \textbf{Training} & \textbf{Dataset} \\
\midrule
Diffusion &
UNet &
\makecell[l]{$N=480$ \\ $\eta = 0.007$ \\ $\omega=1$, $\omega_s=2$\\$\omega_v=0.3$\\$\beta=1.6$ } &
\makecell[l]{$\lambda_{\text{KID}}= 0.05$\\$N_{\text{KID}}=15$\\$\text{ema}=0.999$\\$\text{lr}_{\text{pre}}=10^{-4}$\\$\text{lr}_{\text{fine}}=10^{-5}$} &
\makecell[l]{EchoNet-LVH\\(pretrain)\\DehazingEcho2025 \\(clean subset, finetune)} \\
\midrule
Segmentation &
DeepLabV3+ &
\makecell[l]{$\theta=0.176$ \\ $\sigma_\text{blur}=4.2$} &
\makecell[l]{$\text{lr}=5e-4$} &
\makecell[l]{DehazingEcho2025 \\(noisy subset w/ ROIs)}\\
\bottomrule
\end{tabular}
\end{table}

\begin{figure}
    \centering
    \includegraphics[width=1\linewidth]{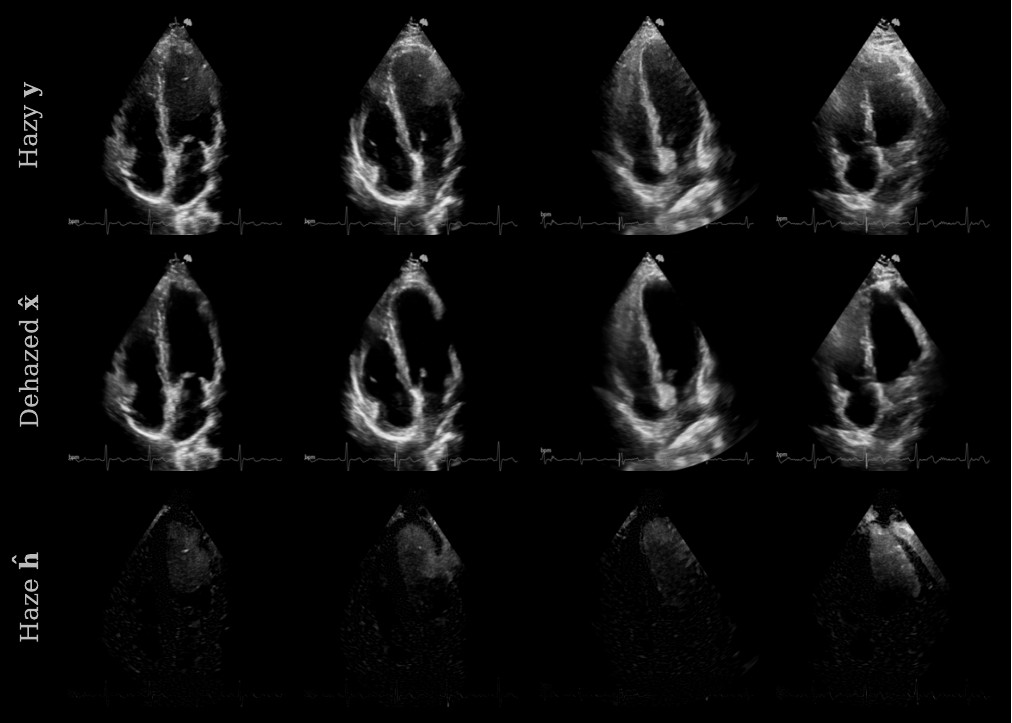}
    \caption{Hazy echocardiographic images $\rvy$ and their decomposition into dehaze doutputs $\hat{\rvx}$ and haze estimates $\hat{\rvh}$ as a result of the submitted algorithm.}
    \label{fig:results}
\end{figure}

\begin{figure}
    \centering
    \includegraphics[width=1.0\linewidth]{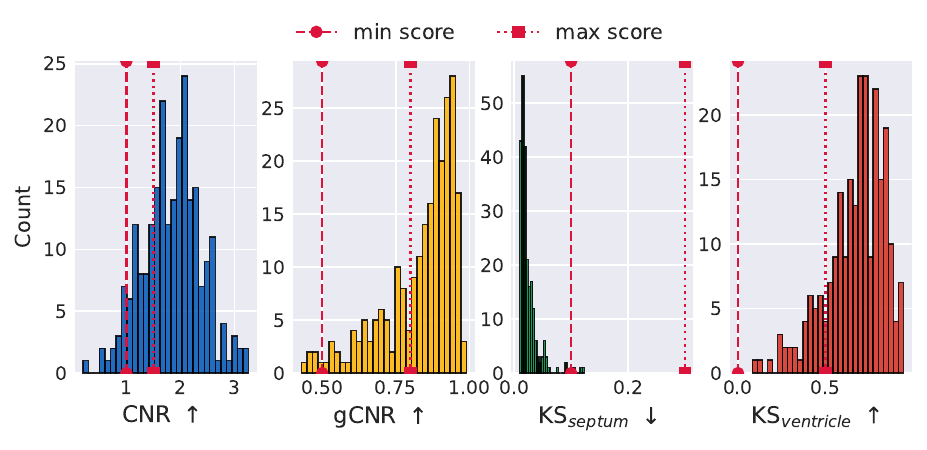}
    \caption{Contrast metrics and KS statistics for the dehazed results from the submitted algorithm. Minimum and maximum obtainable scores as set by the challenge organizers are marked in red.}
    \label{fig:metrics}
\end{figure}

\section{Results and discussion}
A subset of samples from the DehazingEcho2025 noisy set, and corresponding dehazed outputs and haze estimates is shown in Fig.~\ref{fig:results}. A qualitative analysis of the results is shown in Fig.~\ref{fig:metrics}. Both contrast metrics and KS statistics values are plotted for the 237 images in the noisy set that have corresponding ROI masks. The reported \acrshort{fid} obtained under these settings is 61.3. 

One interesting observation is that the hyperparameters yielding the highest challenge score did not necessarily produce the best visual quality, suggesting a misalignment between the evaluation metrics and perceptual fidelity of the dehazed images. For instance, the metrics appear to incentivize an almost binary contrast between the ventricle and septum, which, while improving numerical scores, leads to a loss of subtle structural nuances. This overly sharp separation diminishes the natural appearance of the tissue, which clinicians may find misleading or diagnostically unhelpful. Rather than merely removing haze, the desired outcome is to reveal underlying tissue structures that were previously occluded or obscured. While \acrshort{fid} partially captures this goal, future work should explore more perceptually and clinically aligned evaluation metrics that better reflect meaningful dehazing of echocardiographic images. In the current approach, the ventricle guidance parameter $\omega_v$ and haze prior weighting parameter $\eta$ can be increased and decreased, respectively, to reduce the amount of dehazing in the left ventricle area. 

A final observation is that haze reduction is primarily focused on the left ventricle, driven by both the challenge metrics, centered on septum–ventricle contrast, and the available labels, which include ROIs only for these two regions. As a result, areas like the right ventricle receive less attention. Extending the segmentation model to include the right ventricle would allow for more comprehensive dehazing.

\section{Conclusion}
We present a diffusion-based dehazing algorithm guided by semantic segmentation maps that adaptively modulate the influence of hazy measurements during posterior sampling. Developed in the context of the MICCAI DehazingEcho2025 challenge, our method achieves strong performance across the contrast and fidelity metrics. Preliminary results suggest a potential gap between the existing challenge metrics and perceptual quality, particularly in preserving subtle anatomical details. Nonetheless, the challenge has provided a valuable platform to advance and benchmark dehazing approaches. In particular, our generative modeling approach is able to effectively reduce haze by leveraging semantic guidance to preserve anatomical details. Future work can focus on improved semantic segmentation and development of evaluation metrics that better capture perceptual quality and clinical relevance.
\bibliographystyle{splncs04}
\bibliography{references}

\end{document}